\documentclass[10pt,journal,compsoc]{IEEEtran}
\ifCLASSOPTIONcompsoc
  \usepackage[nocompress]{cite}
\else
  \usepackage{cite}
\fi
\ifCLASSINFOpdf
  \usepackage[pdftex]{graphicx}
\else
\fi
\usepackage{url}

\usepackage{amsmath}
\usepackage{subcaption}
\usepackage{xcolor}

\renewcommand{\eqref}[1]{Eq.~(\ref{#1})}
\newcommand{\figref}[1]{Fig.~\ref{#1}}

\begin{document}
%
\title{Multilayer flows in molecular networks identify biological modules in the human proteome}

%
%
%
%

\author{Giuseppe~Mangioni$^{1,\dag}$, Giuseppe~Jurman$^{2,\star}$, and~Manlio~De Domenico$^{2,\ast}$,\\
$^{1}$~Universit\'{a} degli Studi di Catania, V.le A. Doria 6, 95125, Catania, Italy \\
$^{2}$~Fondazione Bruno Kessler, Via Sommarive 18, 38123 Povo (TN), Italy
\thanks{$^{\dag}$~giuseppe.mangioni@dieei.unict.it, $^{\star}$~jurman@fbk.eu, $^{\ast}$~mdedomenico@fbk.eu}}

%
%

\markboth{}
{}
%



\IEEEtitleabstractindextext{%
\begin{abstract}
A variety of complex systems exhibit different types of relationships
    simultaneously that can be modeled by multiplex networks. A typical problem
    is to determine the community structure of such systems that, in general,
    depend on one or more parameters to be tuned. In this study we propose one
    measure, grounded on information theory, to find the optimal value of
    the relax rate characterizing Multiplex Infomap, the generalization of the
    Infomap algorithm to the realm of multilayer networks. We evaluate our
    methodology on synthetic networks, to show that the
    most representative community structure can be reliably identified when the most
    appropriate relax rate is used. \\
    Capitalizing on these results, we use this measure to identify the most reliable
    meso-scale functional organization in the human protein-protein interaction multiplex network and compare the observed clusters against a collection of independently annotated gene sets from the Molecular Signatures Database (MSigDB). Our analysis reveals that modules obtained with the optimal value of the relax rate are biologically significant and, remarkably, with higher functional content than the ones obtained from the aggregate representation of the human proteome.
    Our framework allows us to characterize the meso-scale structure of those multilayer systems whose layers are not explicitly interconnected each other -- as in the case of edge-colored
    models -- the ones describing most biological networks, from proteomes to connectomes.
\end{abstract}	
}


\maketitle


%
\IEEEpeerreviewmaketitle

\IEEEraisesectionheading{\section{Introduction}\label{sec:introduction}}

%
%
%
%
\IEEEPARstart{A}{} variety of natural and artificial systems can be described in terms of
interactions and relationships among their constituents, which find a suitable
representation in terms of complex networks~\cite{boccaletti2006complex}. 

Despite their wide applicability, single-layer complex networks are not able to
capture the complexity of many empirical systems, as the ones where units
exhibit multiple types of relationships simultaneously. This is the case of
social systems, where an individual can have family, business or trust
interactions with other individuals, or of transportation systems, where
geographical areas might be connected by different transporation means such as
bus, tube, rail, so forth and so on.

\begin{figure}[!ht]
	\centering
    	\includegraphics[width=0.5\textwidth]{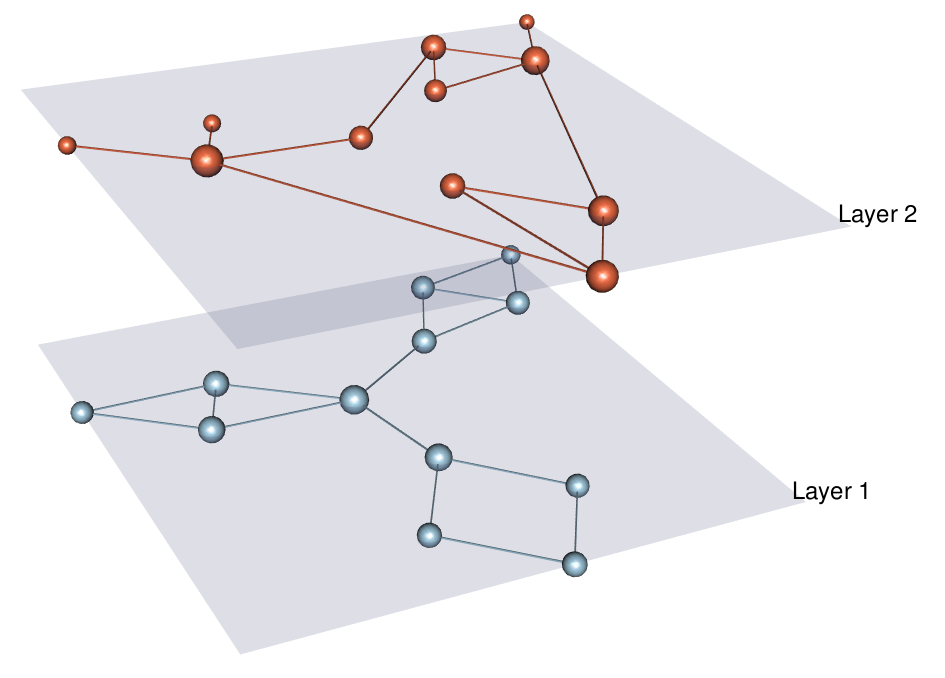}
    \caption{Illustration of a non-interconnected multiplex network (with 2 layers), also called edge-colored multigraph, a special type of multilayer network. This model is defined by i) the existence of a non-empty set of nodes which is common to multiple layers and ii) the absence of explicit information about inter-layer links between node's replicas across layers. The second feature often requires to use a free parameter to modulate the coupling between layers for analytical purposes.}
 \label{fig:multiplex}
\end{figure}

The suitability of multilayer networks for capturing this higher amount of
complexity led to a growing interest in their
study~\cite{kivel2014,Boccaletti20141,de2016physics} and to a more general
mathematical framework, which can be used when nodes are connected to each
other via multiple types of edges or a network changes in
time~\cite{de2013mathematical}. In fact, multilayer networks are more adequate
to model real world interactions that cannot be aggregated into a single
network without a loss, in general, of some important structural or dynamical
properties~\cite{de2015structural,diakonova2016irreducibility} (see Fig.~\ref{fig:multiplex} for an illustration of the type of networks considered in this study).

One of the most attractive problems in network science deals with the
identification of the so-called meso-scale structure of a complex network, a
topic of intensive research activity across multiple
disciplines~\cite{newman2012communities}. Its importance relies in the ability
to unveiling communities of units that, in turn, can be used to explain some
hidden behaviours of networks emerging as the result of the complex interaction
patterns among nodes (or entities).
 
Community detection has been successfully used to analyze the structure of
single-layer networks and for modeling several kinds of interactions, such as
social relationships, genetic interactions among biological molecules or trade
among countries~\cite{girvan2002community,guimera2005functional,palla2005uncovering,palla2007quantifying,bassett2011dynamic,Barigozzi20112051,Barigozzi2011},
just to mention a few (a detailed introduction to communities in networks can
be found in~\cite{Fortunato201075,fortunato2016community}).

In the case of biological systems, community analysis has been used to identify structural and functional modules, in order to determine molecules with similar biological function within a cell and to improve our understanding of life and disease~\cite{barabasi2011network,goh2007human,halu2017multiplex,huttlin2017architecture}. Proteins are molecular building blocks of a cell which play special roles (e.g., catalysis, signalling, etc.) for its modular function and hierarchical organization. In fact, mutation of single genes or altered activation/inhibition regulation quickly propagate to perturbate the protein-protein interaction (PPI) network, causing abnormal functions in tissues and organs that might culminate in diseases. Multilayer network modeling is expected to provide a framework more suitable than traditional aggregated approaches for the analysis of molecular systems and, more specifically of PPI networks. In fact, multilayer networks allow for the integration of multiple information sources, without neglecting or heuristically aggregating different types of interactions among biological units of possibly different type~\cite{de2017integrated}, providing a more realistic framework for systems biology.

Aggregate (also known as ``monoplex'') PPI networks have been successfully used to correlate the cellular function of single proteins with their topological role~\cite{jeong2001lethality}, revealing indispensable proteins, from a network controllability perspective, that turned out to be commonly targeted by disease-causing mutations and human viruses or have been identified as drug targets~\cite{vinayagam2016controllability}. Evidence for a strict relationship between their mesoscale organization and functional segregation within a cell has been provided for both human and non-human organisms~\cite{barabasi2004network,yook2004functional,wu2009integrated}, thus increasing our understanding of functional relationships with genetic disorders~\cite{gandhi2006analysis} and cancer~\cite{rolland2014proteome}.

In this work, we explore the suitability of multilayer community detection for the analysis of PPI interactions, with special focus on the human proteome. The paper is organized as in the following. First, we present the community detection problem in network science and briefly introduce Multiplex Infomap, the methodology used in this work. Second, we introduce the information-theoretic measure -- namely the normalized information loss -- used in this study to select the relax rate, i.e., the parameter characterizing Multiplex Infomap in the analysis of non-interconnected multiplex networks. Third, we present the analysis of synthetic networks to validate the goodness of this measure in determining a suitable relax rate. Finally, we apply the proposed methodology to unravel the meso-scale functional organization of the human PPI multiplex network and we validate the discovered modules against a collection of annotated gene sets. 

\section{Community detection in multilayer networks}\label{sec:community}

Informally, a community is a group of nodes more densely connected each other
inside the group and sparsely connected to nodes outside the group. Despite
this intuitive concept, a precise definition of a community is still a topic of
debate among network scientists.  One of the most adopted formulation is based
on the definition of the so-called null--model, i.e. a model to which the
network can be statistically compared to random expectation in order to assert
the existence of any degree of modularity. Starting from the definition of a
specific null--model -- i.e., a random network satisfying certain requirements,
such as preserving the number of nodes, the number of links and the degree
distribution of the original network -- a modularity function to measure the
quality of a given partition has been introduced in
Ref.~\cite{PhysRevE.69.026113,Newman06062006}. Despite some
limitations~\cite{fortunato2007resolution,good:2010}, the modularity
function has been successfully used as a quality measure to evaluate a given
network partition and as a cost function to be optimized to uncover
communities~\cite{reichardt2004detecting,duch2005community,danon2005comparing}.
Furthermore, the original modularity definition has been extended to
directed~\cite{PhysRevLett.100.118703}, weighted~\cite{PhysRevE.70.056131},
bipartite~\cite{PhysRevE.76.066102} networks and to evaluate partitions with
overlapping communities~\cite{1742-5468-2009-03-P03024}. 

Other than modularity based methods, in literature there exists a lot of
alternative approaches to solve the problem of community
discovering~\cite{Fortunato201075,fortunato2016community}. For example, in
Ref.~\cite{rosvall2008maps} an information-theoretic based method is presented.
This method is based on the formulation of a new quality function called map
equation~\cite{Rosvall2009}, which allows to find the optimal description of the
network by compressing its information flow. The algorithm is the core of {\em
Infomap}\footnote{\url{http://www.mapequation.org/}}, the search method for minimizing
the map equation over possible network partitions.

Many methods and measures developed for single layer networks have been
extended to be applicable to multilayer networks
\cite{de2013mathematical,PhysRevE.89.032804,de2015ranking,cozzo2015structure,%
iacovacci2016functional}. In this context new community detection methods have
been devised, mainly by reusing concepts already developed for single layer
networks. In Ref.~\cite{Mucha876}, the authors proposed a method based on a
generalization of the modularity to multilayer networks. This extended
modularity is mainly based on generalized null models obtained by considering a
Laplacian dynamics~\cite{delvenne2010stability,lambiotte2014random} on the
multilayer network. To compute communities by using such a generalization of
the modularity function, an extension of the Louvain
algorithm~\cite{1742-5468-2008-10-P10008} has been also proposed in
Ref.~\cite{Carchiolo2011}.    

In Ref.~\cite{PhysRevX.5.011027} an extension of the map equation to multilayer
networks is introduced. It is based on the generalization of random walks to
multilayer systems~\cite{de2014navigability}, which in turn are used to
generate the corresponding network flow to be compressed in order to identify
community flows in multilayer networks.  The resulting algorithm -- i.e. {\em
Multiplex Infomap}~\cite{PhysRevX.5.011027} -- is the extension of {\em
Infomap} to the case of multilayer networks.

A drawback of community detection algorithms for
non-interconnected/edge-colored networks -- i.e., systems where inter-layer connectivity is not defined -- is their dependence on at least one
parameter which regulates the structural or dynamical coupling between layers.
In the case of Multiplex Infomap, this parameter is known as the \emph{relax
rate} $r$. The relax rate is the parameter responsible of modeling
movement among layers. At each step of the underlying random walk, there is a $1 - r$
probability that the random walker simply moves to a neighbour in the same
layer, and an $r$ probability that it changes layer, and then moves to a
neighbour on that layer.

To better understand this concept, let us denote by $W_{ij}^{\beta}$ the intra-layer adjacency matrix representing the connectivity of nodes within a generic layer $\beta$ of the multiplex system. Let  $D_{i}^{\alpha\beta}$ represent the inter-layer adjacency matrix of a physical node $i$, encoding the connectivity of that node with its replicas on other layers. Let $S_{i}^{\alpha}~=~\sum_{\beta}D_{i}^{\alpha\beta}$ indicate the inter-layer out-strengths and let $s_{i}^{\beta}~=~\sum_j W_{ij}^{\beta}$ indicate the intra-layer out-strengths of node $i$ in layer $\alpha$ and $\beta$. It follows that the probability of a random walker to move, in general, from node $i$ in layer $\alpha$ to node $j$ in layer $\beta$, is given by~\cite{PhysRevX.5.011027}
\begin{align}
\mathcal{P}_{ij}^{\alpha\beta}(r) =  (1-r) \delta_{\alpha\beta} \frac{W_{ij}^{\beta}}{s_{i}^{\beta}} + r \frac{W_{ij}^{\beta}}{\sum_{\beta}s_{i}^{\beta}}.\label{eq:multilayertransitionprob}
\end{align}

The choice of $r$ is crucial and, in general, it depends on the network under
analysis. While empirical results suggest that values smaller than $0.5$ are
generally appropriate for most networks~\cite{PhysRevX.5.011027}, finding the
actual optimal value is still an unsolved problem. Moreover, in community
detection the concept of an absolute optimal simply does not exist, as it is
difficult to ascertain whether the chosen algorithm is able to detect the
absolute optimal partition. In fact, a safer approach is to assess that a
certain partition can be optimal with respect to a specific algorithm. In this
study, our goal is to find the value of $r$ which provides the best possible
partition with respect to \emph{Multiplex InfoMap} in the case of multilayer
systems where the strength of coupling among layers is unknown.
As thoroughly explained in \cite{Peele1602548}, the problem of finding the best
possible partition in community detection is ill-defined -- in the case
of multiplex networks and, more generally, in complex networks -- and
the result of the proposed procedure will provide optimal partitions
with respect to the information-theoretic quality functions used in this
work.

\section{Information-theoretic approach to parameter selection}
\label{sec:approach}

Multiplex InfoMap is an algorithm which optimizes the map
equation~\cite{rosvall2008maps}, a measure of the information-theoretic duality
between data compression and the problem of extracting significant information
from compressed data. Given that its roots lie firmly in the realm of
information theory, it is natural to develop an information-theoretic algorithm
to determine the relax rate producing an optimal partition, with respect to
some criteria. In the domain of information theory, this partition would be the
one which retains the most information about the network inside the
communities. In literature, there are several attempts at exploiting the
concepts of information theory to evaluate the quality of a partition~\cite{PhysRevE.71.046117,Rosvall01052007,rosvall2008maps,PhysRevLett.110.148701}. Here,
we investigate the suitability of information loss~\cite{Rosvall01052007} for optimal selection of relax rate.

Information loss occurs when a certain source of information is compressed in a
way that some of the information is discarded as a result of the compression.
Since any source of information can be fed to compression algorithms, we are
now going to describe how to compress a network, and the information loss that
derives from this operation.

Compressing a network $X$ involves finding some representation $Y$ that only
keeps part of the available information on connectivity. 
In the following, we indicate by $H(X)$ the information entropy of the
random variable $X$ encoding the original network connectivity, while we
indicate by $H(Y)$ the information entropy of the random variable $Y$,
which provides a simplified and coarse-grained representation of the
system. As we will see in the following, an explicit calculation of
$H(X)$ and $H(Y)$ is not required.
In \cite{Rosvall01052007}, the authors compress the network into a representation
that preserves the information contained inside the communities in order to
evaluate how much information is required to rebuild $X$ given the
representation $Y$. If we name this quantity $H(X|Y)$, given that $H(X)$ is the
average amount of information required to describe $X$, we can compute it from
mutual information $I(X;Y)$ between $X$ and $Y$ by
\begin{eqnarray}
\label{eq:Hcond}
    H(X|Y) = H(X) - I(X;Y).
\end{eqnarray}
The compressed representation $Y$ is still a graph where each node is a
community and links between nodes are inter-community connections. Hence, we
may completely describe $Y$ with the tuple $(N, L)$, with $N = \{n_{i}\}$,
where $n_i$ is the number of nodes of the $i$-th community, and $L =
\{l_{ij}\}$, where $l_{ij}$ is the number of links that go from community $i$
to community $j$. Note that the definition provided is exactly equivalent to
the cross-entropy (or negative log-likelihood) of the stochastic block model
(SBM) \cite{peixoto:2012}, a widely adopted generative model for random graphs. If we assume
that there are $m$ communities, in the simplest case of undirected and
unweighted networks, Eq.\,(\ref{eq:Hcond}) reduces to:
\begin{eqnarray}
    H(X|Y) = \log_2\left[ \prod_{i=1}^m \binom{n_i (n_i - 1) / 2}{l_{ii}} 
                    \prod_{j<i} \binom{n_i n_j}{l_{ij}} \right].
\end{eqnarray}
This formula accounts for all the possible ways to arrange the links that go
from nodes of community $i$ to nodes of community $j$, hence representing all
the possible configurations of networks that can be reconstructed knowing $Y$.
The higher the value, the more information is contained in the inter-community
links. Extending this formula for directed networks is straightforward, since
one should evaluate the possibility that a link can connect two nodes in two
different ways (from $i$ to $j$ and vice-versa):
\begin{equation}
    \label{eq:unweighted-hz}
    H(X|Y) = \log_2\left[ \prod_{i=1}^m \binom{n_i (n_i - 1)}{l_{ii}} 
                    \prod_{j\ne i} \binom{n_i n_j}{l_{ij}} \right].
\end{equation}

\begin{figure*}[!ht]
    \centering
    \begin{subfigure}[b]{0.23\textwidth}
        \includegraphics[width=\textwidth]{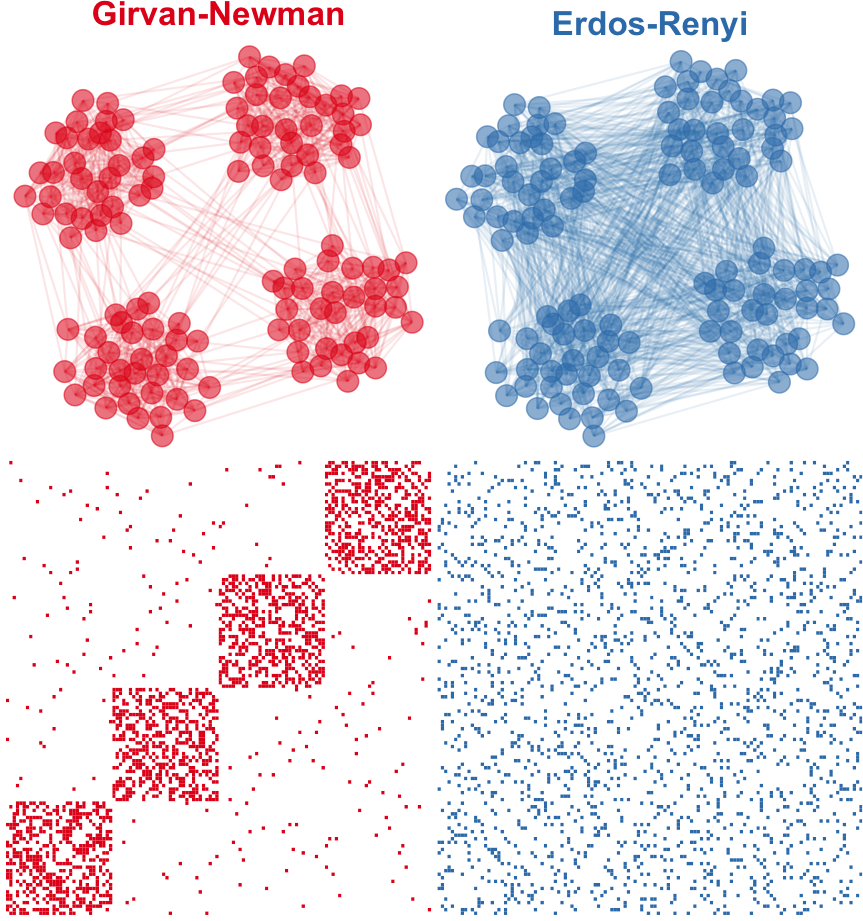}
        \caption{GN/ER}
        \label{fig:GNER}
    \end{subfigure}
    \begin{subfigure}[b]{0.23\textwidth}
        \includegraphics[width=\textwidth]{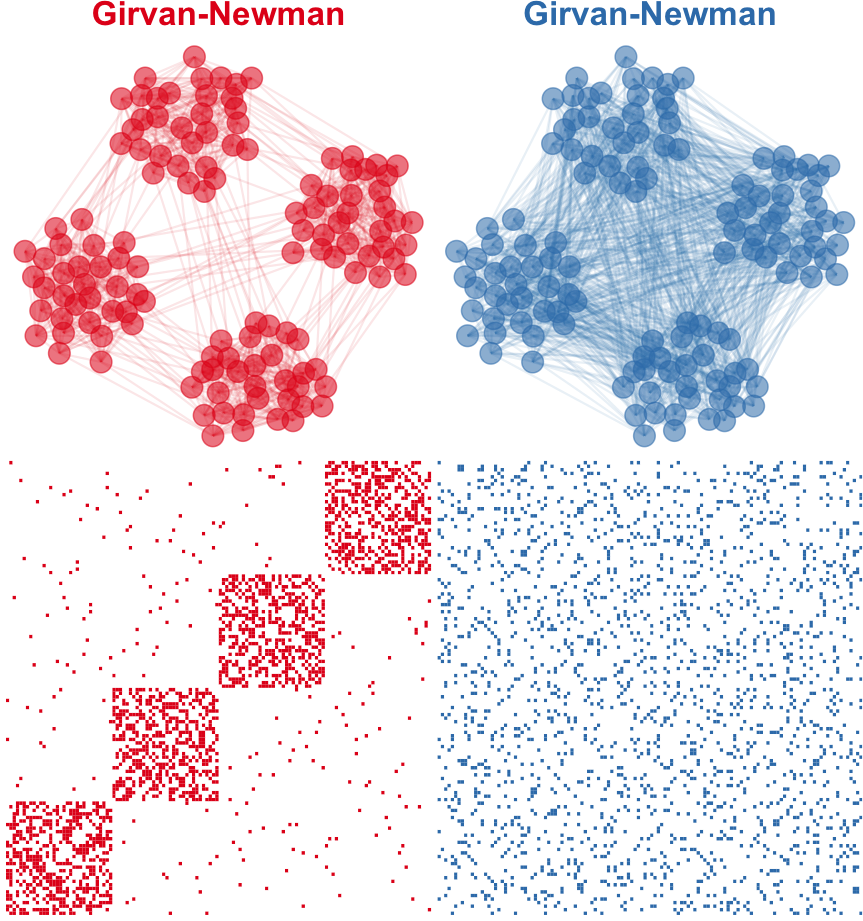}
        \caption{GN/GN 10\%}
        \label{fig:GNGN10}
    \end{subfigure}
    \begin{subfigure}[b]{0.23\textwidth}
        \includegraphics[width=\textwidth]{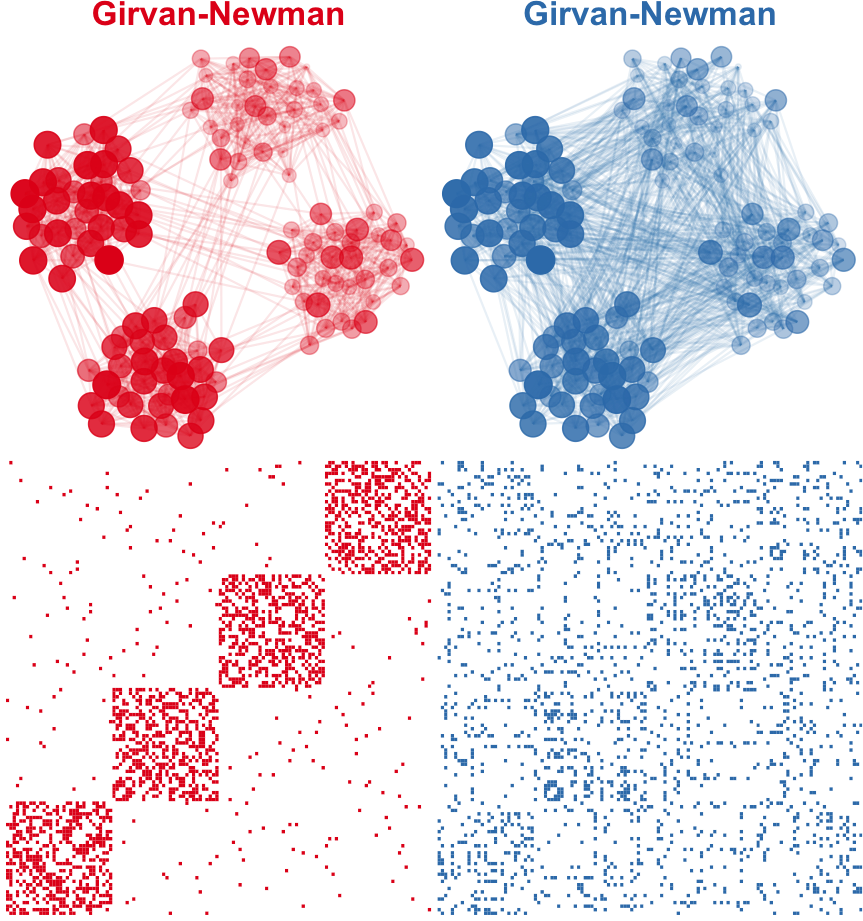}
        \caption{GN/GN 50\%}
        \label{fig:GNGN50}
    \end{subfigure}
    \begin{subfigure}[b]{0.23\textwidth}
        \includegraphics[width=\textwidth]{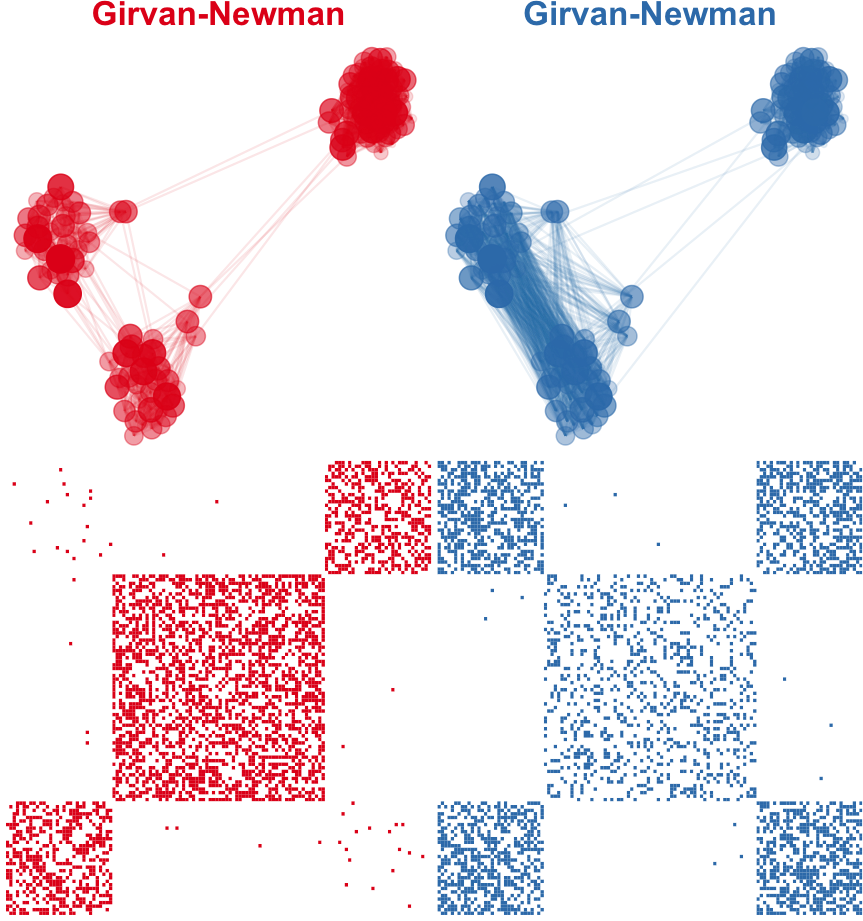}
        \caption{GN/cGN}
        \label{fig:GNCGN}
    \end{subfigure}
    \caption{Synthetic multiplex network models considered in this study as benchmarks. Network of 128 nodes and different community structures across layers are shown. (a) A layer with a strong community structure (Girvan-Newman model) coupled to a layer with a homogeneous structure (Erdos-Renyi model) playing the role of structural noise. (b) Two Girvan-Newman networks with 10\% of nodes belonging to overlapping communities across layers. (c) As in (b), but with 50\% overlapping community structure. (d) A layer with a network generated by using the Girvan-Newman model (GN) coupled to a layer with the {\em complementary} network (cGN). See the main text for further details.}
    \label{fig:syn}
\end{figure*}

Weights can be included as well, to account for more complex structures. For
each link $l_{ij}$ of $Y$, which represents the total number of links from
community $i$ to community $j$, we have a quantity $w_{ij}$ encoding the sum of
the weights of links that go from $i$ to $j$. Ideally, each configuration
reconstructed from $Y$ using \eqref{eq:unweighted-hz} generates further
configurations if we consider all the possible ways to distribute $w_{ij}$
among $l_{ij}$ links.  The number of those configurations is infinite if the weights
are real numbers: given a weight $w_{ij}$, the problem is analogous to
splitting the interval $[0, w_{ij}]$ in $l_{ij}$ parts, and since any real
interval is uncountable, there are infinite ways to make the partition.
However, if we impose the restriction that the weights are natural numbers, the
number of partitions can be calculated as follows. First, we assign the weight
$1$ to each one of the $l_{ij}$ links, thus imposing the restriction $w_{ij}
\ge l_{ij}$. Since we already distributed $l_{ij}$ out of the total $w_{ij}$,
calculating all possible distributions of the remaining $w_{ik} - l_{ij}$ among
$l_{ij}$ links depends on combinations with replacement:
\begin{eqnarray}
    C^R(l_{ij}, w_{ij} - l_{ij}) 
    &=& \frac{(l_{ij} + w_{ij} - l_{ij} - 1)!}
        {(w_{ij} - l_{ij})!(l_{ij} - 1)!}\nonumber\\
    &=& \frac{(w_{ij} - 1)!}{(w_{ij} - l_{ij})!(l_{ij} - 1)!}\nonumber\\
    &=& \binom{w_{ij} - 1}{l_{ij} - 1}.
\end{eqnarray}
Thus, we can update equation \eqref{eq:unweighted-hz} to include all the
possible ways to distribute $w_{ij}$ among $l_{ij}$ links:
\begin{eqnarray}
    \label{eq:weighted-hz}
    H(X|Y) &=& \log_{2}\left[ \prod_{i=1}^m \binom{n_i (n_i - 1)}{l_{ii}}
        \binom{w_{ii} - 1}{l_{ii} - 1} \times \right.\nonumber\\
        &&\left.\times \prod_{i\ne j} \binom{n_i n_j}{l_{ij}} 
        \binom{w_{ij} - 1}{l_{ij} - 1} \right].
\end{eqnarray}
A more general formula, accounting for the possibility of self-links (e.g.,
useful for modeling citation networks) is given by
\begin{equation}
    H(X|Y) = \log_{2}\left[ \prod_{i=1}^m \prod_{j=1}^m \binom{n_i n_j}{l_{ij}} 
        \binom{w_{ij} - 1}{l_{ij} - 1} \right].
\end{equation}
Since $H(X|Y)$ represents the information that is lost when compressing the
network, a good compression requires $H(X|Y)$ to be as small as possible:
hence, our goal is to minimize this quantity.
It is worth remarking that it is possible to learn the latent block
structure in presence of real-valued weights through the use of a
parametric distribution, as shown in \cite{doi:10.1093/comnet/cnu026}, by exploiting the fact that
the proposed measure can be interpreted as a log-likelihood of a SBM.

For practical applications, it might be useful to define a standardized version of this measure. If we name the information loss $H_r(X|Y)$ for a certain relax rate $r$, we define its normalized version as:
\begin{equation} \label{eq:loss-norm}
   H_{r}^{*}(X|Y) = \frac{H_{r}(X|Y) - \min\limits_{0 < r \leq 1} H_{r}(X|Y)}{\max\limits_{0 < r \leq 1} H_r(X|Y) -
       \min\limits_{0 < r \leq 1} H_r(X|Y)}.
\end{equation}

\section{Analysis of synthetic network models}
\label{sec:synthetic}

To better understand the suitability and the limitations of the the proposed measure, we analyze a set of synthetic benchmark networks. The multiplex toy models consist of nodes which are connected in different ways on two layers, while inter-layer connectivity is not given explicitly. We consider four kinds of benchmarks:

\begin{figure*}[!t]
    \centering
    \begin{subfigure}[b]{0.4\textwidth}
        \includegraphics[width=\textwidth]{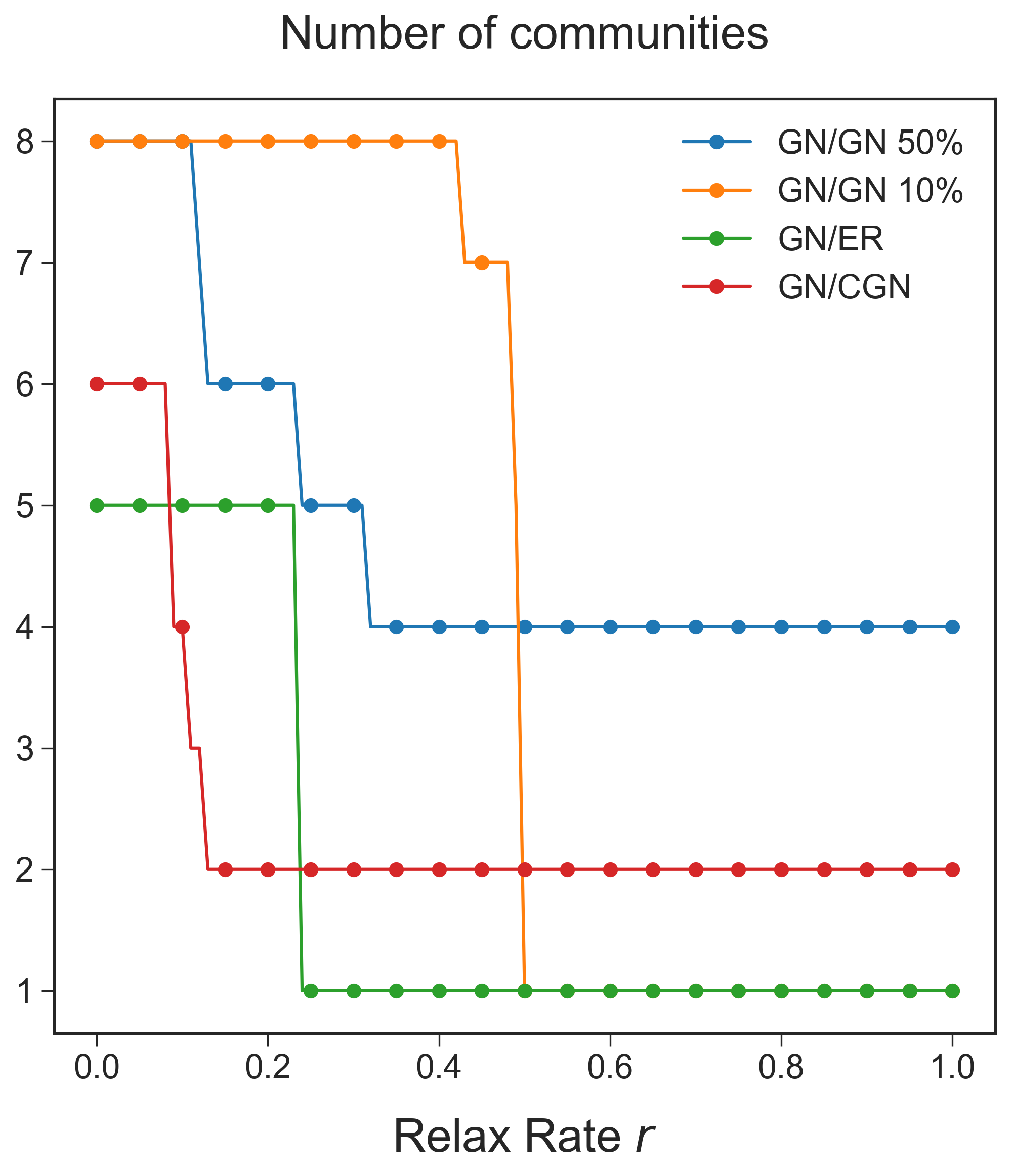}
        \caption{}
        \label{fig:comms-syn}
    \end{subfigure}
    \begin{subfigure}[b]{0.41\textwidth}
        \includegraphics[width=\textwidth]{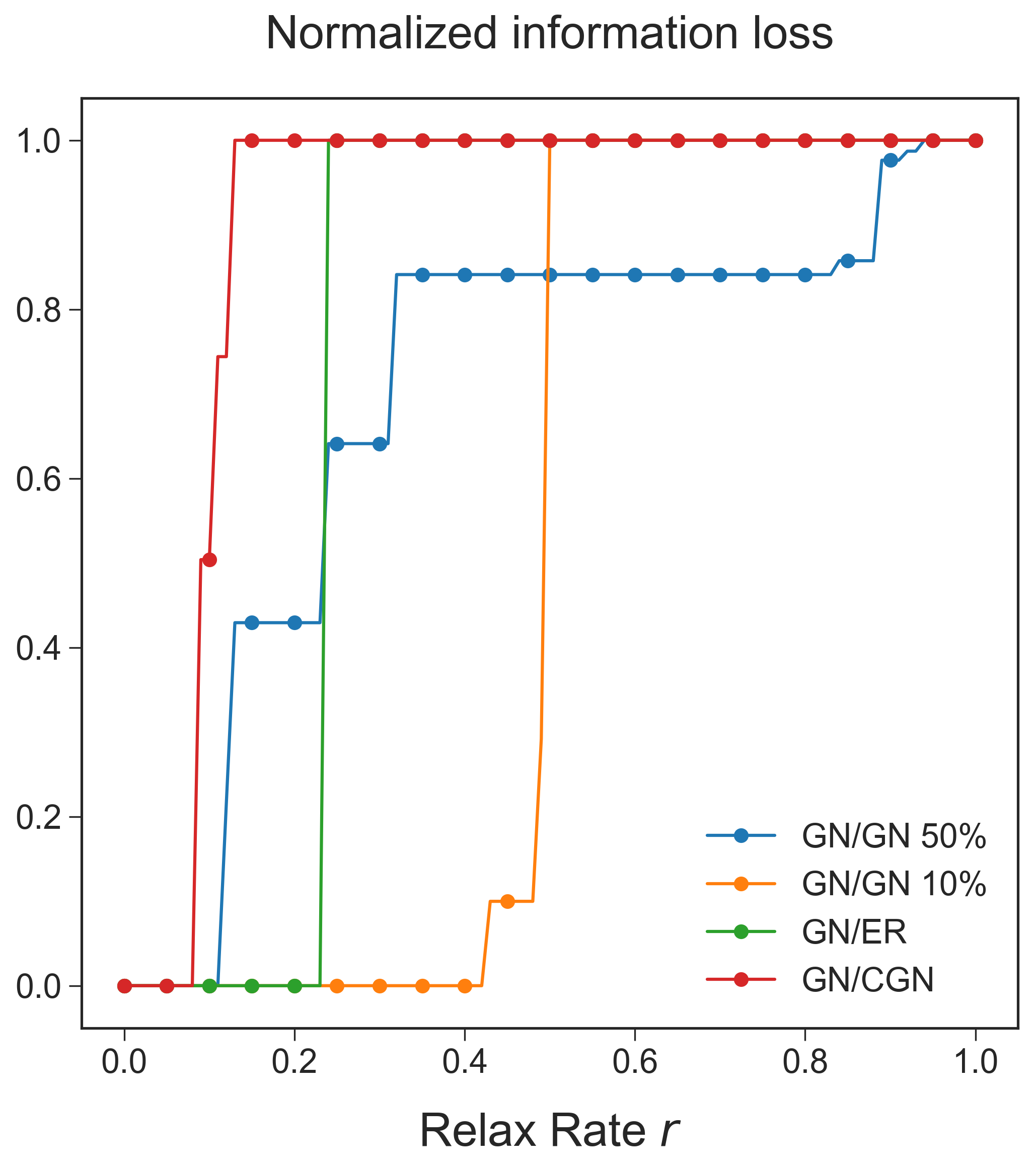}
        \caption{}
        \label{fig:loss-syn}
    \end{subfigure}
    \caption{(a) Number of communities and (b) normalized information loss, for different synthetic network models, while varying the value of the relax rate.}
    \label{fig:measures-syn}
\end{figure*}

\begin{itemize}
    \item \textbf{GN/ER.} This system is generated by combining an Erdos-Renyi layer
        with one generated by the Girvan-Newman (GN)
        benchmark~\cite{girvan2002community}. A community structure is only present in the GN
        layer, hence this multiplex network is used to test the impact of coupling noise to a structured population.
    \item \textbf{GN/GN.} This system consists of two
        GN networks, with tunable cross-layer
        community overlap, generated as following. First, we create a
        single-layered GN network and duplicate it to generate a multiplex network with two layers.  Then, in second layer, we iteratively swap the neighbours of pairs of randomly selected nodes to change the community structure. The swapping procedure is repeated until
        the ratio of community overlapping across layers, defined by the
        fraction of nodes that belong to the same communities across layers, is reached. Two classes of networks are generated, corresponding to a different amount of overlapping
        across layers (50\% and 10\%, respectively).
    \item \textbf{GN/CGN.} This system consists of two layers, one GN network and one complementary GN (CGN). In order to explain how these two layers are generated, let's suppose that communities in each layer are labeled with the integers $\{1,2,3,4\}$. In the first layer, we preserve the structure of communities $1$ and $2$, while nodes in communities $3$ and $4$ are randomly rewired with a given probability $p=0.5$. A fraction of $10\%$ of nodes in communities $3$ and $4$ are then connected to nodes in communities $1$ and $2$ chosen at random. The same algorithm is then applied to layer 2, but preserving communities $3$ and $4$ and randomly rewiring nodes in communities $1$ and $2$.  
\end{itemize}

These synthetic network models (see Fig.~\ref{fig:syn}) are used to test the ability of our methodology to detect the most relevant community structure among the ones identified by Multiplex Infomap for varying relax rates. Results are shown in \figref{fig:measures-syn}. Our analysis highlights some special properties of normalized information loss when an underlying community structure is present or absent. First, it is worth remarking that transitions between two or more regimes are always observed: in fact, the relax rate acts as a dynamical multiresolution parameter, allowing to identify communities at different dynamical scales. Let us consider, for instance, the GN/ER model in Fig.~\ref{fig:syn}a: for $r$ smaller than 0.25 we identify 5 communities (the four planted partitions in the GN layer plus the ER layer, acting as a single community), whereas a sharp transition towards 1 community is observed above that value. In this latter case, the random walkers are exploring the ER layer too often and the relevance of the strong community structure planted in the GN is washed out. While this behavior finds a clear explanation, it might be useful for applications to identify a dynamical scale, by means of a suitable value of the relax rate, where the identified meso-scale organization is more representative of the system. Here, the normalized information loss we have previously introduced plays a crucial role: in fact, it is zero before the transition point and explosively increases to one above it (see Fig.~\ref{fig:syn}b). Therefore, our measure is highlighting that, from an information-theoretic perspective, the meso-scale found below the transition point provides more information about the system.

Similar arguments can be given to explain the behavior of transitions in the other network models considered in the same figure. The transition point can change, depending on the model and its complexity, and we might observe more than a single sharp transition point, each one identifying a transition between different dynamical regimes. Remarkably, in all cases the evolution of normalized information loss resembles the evolution of the number of communities while providing, at the same time, a quantitative measure which helps to identify the range of dynamics describing the system with minimum information loss.

\begin{figure*}[ht]
    \centering
    \includegraphics[width=\textwidth]{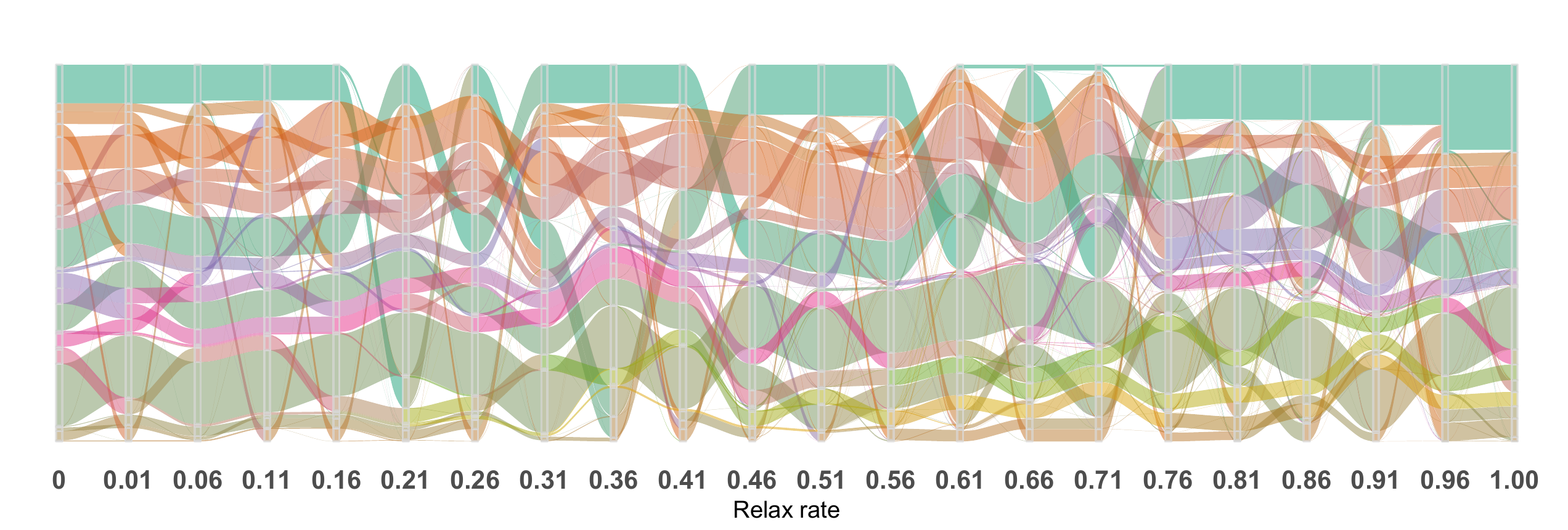}
    \caption{Changes in the meso-scale organization of human multiplex proteome for varying relax rate. Clusters with at least 100 proteins are considered for clarity. This alluvial plot shows how partitions split and merge for increasing rate: larger clusters are quite stable, highlighting that differences in meso-scale are mostly due to smaller sets of proteins.}
    \label{fig:clu-evo}
\end{figure*}

\section{Analysis of the \emph{Homo Sapiens} PPI Multiplex Network} \label{sec:empirical}

In agreement with previous studies~\cite{muxviz,de2015structural}, we analyze multiplex molecular networks built from the BioGRID (Biological General Repository for Interaction Datasets~\cite{stark2006biogrid}) a public database with data compiled through comprehensive curation efforts, that stores and disseminates genetic and protein interaction information about simple organisms and humans\footnote{\url{http://thebiogrid.org}}). The curated set used in this study (BioGRID~3.4.155, updated to December~2017) consists of physical and genetic relations including interactions, chemical associations, and post-translational modifications (PTM) from 63,959 publications, leading to 1,168,521 non-redundant interactions, 1,507,991 raw interactions, 11,820 non-redundant chemical associations, 27,785 raw chemical associations, 19,981 Unique PTM Sites, and 18,578 Un-Assigned PTMs. 

More specifically, we focussed our attention on \emph{homo sapiens}. In this data set, the networks represent PPI of two types, namely genetic and physical, and the layers correspond to a finer classification into seven types of interaction of different nature, i.e., physical, direct, co-localization, association, and suppressive, additive or synthetic genetic interaction. The total number of proteins in the network is 21,591, linked by 338,113 interactions across all layers. 

To compare our results against more traditional approaches based on the study of the aggregate representation of the same system, we build the corresponding aggregate network which consists of 315,766 unique weighted interactions. It is worth noting that 6.6\% of the total number of interactions are present in at least two different layers.

\begin{figure}[ht]
    \centering
    \includegraphics[width=0.49\textwidth]{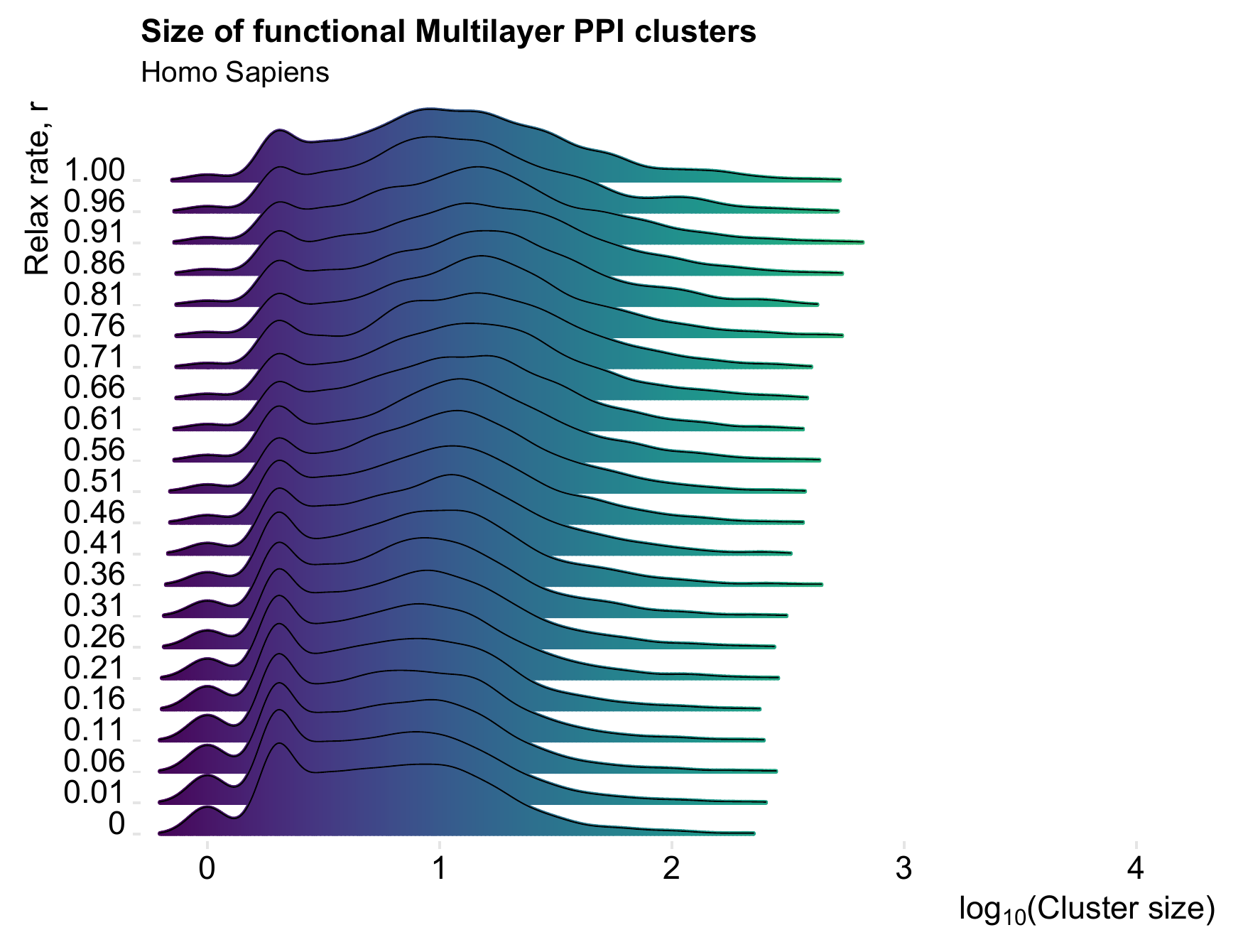}
    \caption{Changes in the meso-scale organization of human multiplex proteome: the distribution of clusters' size is shown for varying relax rate. For increasing relax rate, the number of very small clusters (size smaller than 2) decreases while larger functional clusters of proteins tend to form.}
    \label{fig:clu-distr}
\end{figure}

First, we identify the meso-scale organization of the multiplex representation of the human proteome for varying relax rate. In Fig.~\ref{fig:clu-evo}, we show how the identified communities split and merge for different values of $r$: larger clusters are quite stable, highlighting that differences in meso-scale are mostly due to smaller sets of proteins. Figure~\ref{fig:clu-distr} shows how the distribution of clusters' size changes depending on relax rate, providing a different analysis of the human proteome meso-scale organization. The number of very small clusters (size smaller than 2) decreases whereas larger functional clusters tend to form for increasing relax rate.

Both the multiplex and the aggregate representations are then tested for the biological functional content of their functional clusters through a standard enrichment-based strategy using the Molecular Signatures Database\footnote{\url{http://software.broadinstitute.org/gsea/msigdb/collections.jsp}} (MSigDB)~\cite{subramanian05gene,liberzon11molecular,liberzon15molecular}. In the current version 6.1 (Oct 2017), MSgiDB includes 17,786 gene sets divided into 8 major collections to be used as the knowledge base for enrichement studies. To obtain a quantitative assessement of the functional content associated to the sets of clusters identified from the multiplex network -- for varying relax rate -- or its aggregate representation, we adopt the following procedure: given a set of clusters $\mathcal{C}$ and the MSigDB $\mathcal{M}$, we compute the total functional content as the cumulative fraction of genes shared between the clusters and the collections of annotated gene sets:
\begin{displaymath}
\mathrm{TFC}(\mathcal{C}) = \sum_{\substack{C\in\mathcal{C},|C|\geq 10\\ M\in\mathcal{M}}} J(C,M) 
= \sum_{\substack{C\in\mathcal{C},|C|\geq 10\\ M\in\mathcal{M}}}\frac{C \cap M}{C \cup M}\ ,
\end{displaymath}
where $J(\cdot, \cdot)$ is the Jaccard index~\cite{jaccard02lois,jaccard08nouvelles}. It is worth noting that in our analysis we have considered only clusters with at least 10 genes to allow for statistically meaningful enrichment analysis.

\begin{figure}[ht]
    \centering
    \includegraphics[width=0.49\textwidth]{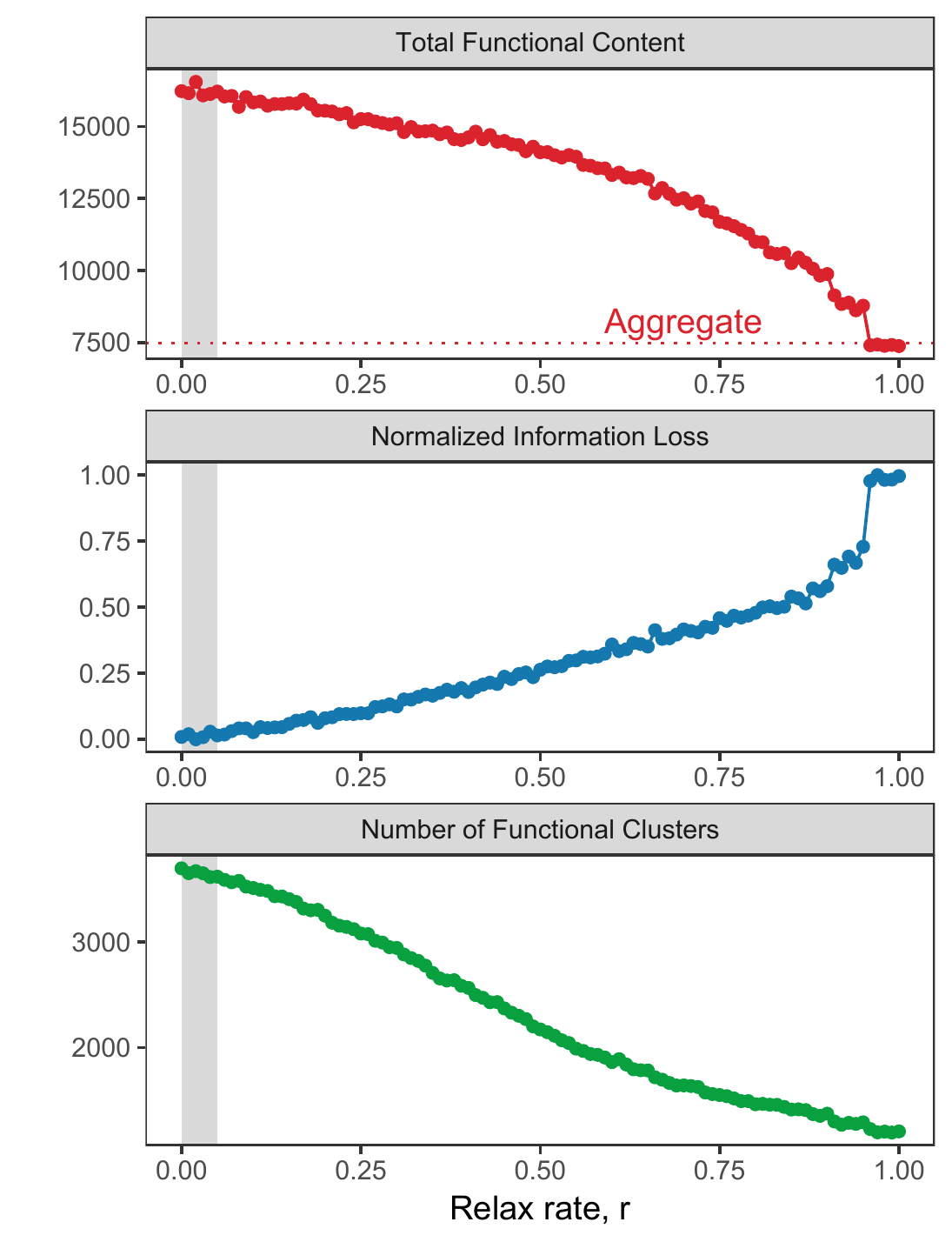}
    \caption{Analysis of the meso-scale organization of genes in the human proteome. \emph{Top panel:} total functional content of the cluster collections as a function of the relax rate, computed as the sum of the Jaccard indices of the intersections between each cluster and each gene set in MSigDB (see the text for details). The content is monotonically decreasing for increasing relax rate: for $r\approx 1$, the functional content coincides with the functional content of the cluster collection obtained from the aggregate representation of the system. \emph{Middle panel:} normalized information loss (Eq.~(\ref{eq:loss-norm})) for increasing relax rate, highlighting an increasing degradation of the information content. \emph{Bottom panel:} decreasing number of identified functional clusters for increasing relax rate. In all panels, the shaded area highlights the range of relax rates where normalized information loss is minimum which, remarkably, coincides with the range where the biological enrichment of identified clusters exhibits maximum functional content.}
    \label{fig:enrich}
\end{figure}

The plot of the total functional content as a function of the relax rate is shown in the top panel of Fig.~\ref{fig:enrich}: the functional content of the set of clusters is decreasing for increasing value of the relax rate, and for relax rates close to one, the content coincide with the functional content of the clusters obtained in the aggregate case. Overall then the largest biological meaningfulness is reached in the multiplex case, for relax rates close to zero, i.e., when random walkers exploring the multiplex systems do not switch layer frequently. This results is remarkable because in perfect agreement with the range of $r$ where normalized information loss is minimum (middle panel of Fig.~\ref{fig:enrich}).

\section{Conclusions}
\label{sec:conclusion}

In this paper, we have proposed an information-theoretic approach for parameter selection in community detection analysis performed with the Multiplex Infomap algorithm, the multilayer variant of the well-known Infomap algorithm. In fact, in the specific case of non-interconnected (i.e., edge-colored) networks, Multiplex Infomap depends on the relax rate, which is responsible for coupling the layers of the network and allows to study the system at multiple dynamical scales.

To tackle the problem of selecting an optimal value for this parameter, our work starts from the observation that the information content of a complex network can be fully encoded by its adjacency matrix representation. In a complex network with a community structure, we can distinguish between information retained within communities, encoded by intra-community links, and information retained among different communities, encoded by inter-community links. Since a good partition of the system is expected to keep within the same community most of the links, or the information flow, we have shown that it is possible to find the most reliable partitions by minimizing the information contained in inter-community connectivity. We achieved this goal by evaluating one information-theoretic measure, namely the normalized information loss.

We have analyzed the behavior of this measure for varying relax rate in both synthetic and empirical networks. Results from trivial benchmarks confirmed our expectations, whereas results from toy models with a non-trivial multiplex community structure show that optimal values of the relax rate are range between $0.1$ and $0.5$, the region of the parameter space where inter-layer coupling -- indirectly caused by random walkers switching among layers -- is strong enough for multilayer effects to become significant.

We have applied the proposed framework to the multiplex human proteome, consisting of proteins interacting physically and genetically. We have identified seven different layers by using the Biological General Repository for Interaction Datasets (BioGRID), in agreement with previous studies. Our analysis of this multiplex PPI network and of its aggregate representation has highlighted the existence of a range of relax rates close to zero where the information content of the meso-scale organization is maximum or, equivalently, where the information loss is minimum. To validate the identified functional clusters, we have performed an independent analysis by performing a biological enrichment through the well known Molecular Signatures Database (MSigDB), a huge collection of manually curated gene sets with known cellular function. Remarkably, the enrichment analysis provides a maximum functional content, which is biologically significant, for the same range of relax rates where normalized information loss is minimum.

Our study provides a quantitative approach for the selection of a suitable value of relax rate in multilayer community detection on non-interconnected multiplex networks. The application of our framework to the study of human proteome provides results which outperform traditional approaches, such as the ones based on the analysis of aggregate representations of multiplex systems. Future applications of our framework include the analysis of different biological systems, from other organisms' multiplex PPI networks to the human brain.

\vspace{6pt} 


\ifCLASSOPTIONcaptionsoff
  \newpage
\fi



\bibliographystyle{IEEEtran}

\bibliography{biblio}

\end{document}